# A unified framework for sensitivity comparison across stimulated Raman, stimulated Raman photothermal, and mid-infrared photothermal microscopy

Keiichiro Toda[1*], and Takuro Ideguchi[1,2,3]

[1] Institute for Photon Science and Technology, The University of Tokyo, Tokyo, Japan

[2] Department of Advanced Materials Science, The University of Tokyo, Chiba, Japan

[3] RIKEN Center for Advanced Photonics, RIKEN, Saitama, Japan

[*] keiichiro--toda@g.ecc.u-tokyo.ac.jp

**Abstract**

Label-free vibrational microscopy based on Raman scattering and mid-infrared (MIR) absorption has advanced biological imaging through the development of diverse high-sensitivity modalities. However, rigorous sensitivity comparisons across these techniques remain lacking. Here, we establish a unified theoretical framework for quantitatively comparing the shot-noise-limited sensitivities of stimulated Raman scattering (SRS), MIR photothermal (MIP), and stimulated Raman photothermal (SRP) microscopy. The framework is built on two quantities: an effective absorption coefficient, which places SRS on the same footing as linear absorption, and a photothermal (PT) factor, which isolates the contribution specific to PT readout. Using experimentally realistic parameters, we show that MIR absorption coefficients for representative polar vibrational modes are typically about two orders of magnitude larger than the effective absorption coefficients of SRS, even for strong off-resonant Raman bands. In MIP, however, this intrinsic advantage is partly offset by a small PT factor, limiting the net sensitivity gain over SRS to several-fold and, at most, about one order of magnitude. In SRP, by contrast, burst-pulse thermal accumulation substantially increases the PT factor. Under ideal shot-noise-limited conditions with matched per-pulse excitation-energy conditions in water, SRP is predicted to provide sensitivities several-fold higher than those of SRS and to approach those of MIP. This framework provides a common basis for sensitivity assessment across vibrational imaging modalities and clarifies how absorption strength, photothermal accumulation, and readout conditions determine the practical advantages of each modality.

## 1. Introduction

Label-free vibrational imaging based on mid-infrared (MIR) absorption and Raman scattering offers a powerful tool for chemically resolved analysis in the life sciences. Spontaneous Raman microscopy can map intracellular chemical composition with sub-micrometer spatial resolution and has been applied to lipidomics, bacterial phenotyping, and metabolic profiling[1-3]. However, the inherently small scattering cross-section of spontaneous Raman signals ($10^{-28}$-$10^{-29}$ cm$^2$) necessitates long acquisition times, often on the order of tens of minutes, which limits its utility for real-time analysis of cellular composition. By contrast, MIR absorption (MIA) microscopy benefits from much larger absorption cross-sections ($10^{-18}$-$10^{-19}$ cm$^2$) and therefore supports substantially faster imaging. This advantage has enabled applications such as metabolic imaging in brain development and tumor progression[4]. Its spatial resolution, however, is fundamentally limited by the diffraction limit at MIR wavelengths (2.5-25 μm), making it unsuitable for single-cell imaging. Although spontaneous Raman microscopy and MIA microscopy offer complementary advantages, their respective limitations have restricted their broader use in biological research.

Coherent Raman scattering (CRS)[5,6] and MIR photothermal (MIP) microscopy[7-10] have emerged as major strategies to overcome the respective limitations of spontaneous Raman and MIA microscopy. To compensate for the weak signal levels of spontaneous Raman scattering, CRS techniques such as coherent anti-Stokes Raman scattering (CARS)[5] and stimulated Raman scattering (SRS)[6] exploit nonlinear optical processes to enhance detection sensitivity by two to three orders of magnitude. As a result, CRS has become a leading platform for high-speed cellular vibrational imaging, enabling video-rate two-dimensional imaging[11,12] and high-speed hyperspectral acquisition[13,14]. By contrast, MIP microscopy addresses the spatial-resolution limitation of MIA microscopy. By detecting photothermal refractive index (RI) changes induced by MIA with visible (VIS) probe light, it provides MIR chemical contrast at the diffraction-limited resolution of VIS wavelengths. Recent improvements in detection sensitivity have extended MIP to both two-dimensional and three-dimensional imaging at video rate[15-17]. More recently, SRS photothermal (SRP) microscopy[18] has emerged as a hybrid strategy that combines CRS excitation with photothermal readout and has achieved sensitivities exceeding those of conventional CRS. These advances have given rise to a wide range of highly sensitive vibrational imaging modalities, making rigorous cross-platform sensitivity comparisons increasingly important. However, such a comparison has remained elusive. A unified quantitative framework for evaluating the sensitivities of CRS, MIP, and SRP is still lacking, largely because these modalities are governed by fundamentally different signal-generation and detection principles.

In this work, we establish a unified theoretical framework for quantitatively comparing the sensitivities of CRS, MIP, and SRP microscopy. To place these modalities on a common footing, we introduce an effective absorption coefficient derived from the cross-sections relevant to SRS and MIA, together with a photothermal (PT) factor that captures the contributions specific to photothermal readout. This formulation enables rigorous sensitivity analysis under shot-noise-limited conditions across modalities governed by different physical principles, thereby providing a general basis for assessing detection limits and guiding the design of next-generation vibrational imaging methods.

## 2. Principles

### Direct vibrational detection by MIA and CRS microscopy

We first introduce the principles of MIA and CRS microscopy that provide the foundation for the present sensitivity comparison. Figure 1a shows the energy-level diagrams of MIA and SRS. We use SRS as the representative CRS modality because SRS and CARS are predicted to have comparable shot-noise-limited sensitivities[19]. The conclusions drawn here therefore remain broadly applicable across CRS modalities. In MIA microscopy, incident MIR light in the 2.5-25 μm range resonantly excites molecular vibrations, and image contrast is generated by attenuation of the transmitted MIR intensity. In SRS microscopy, by contrast, co-axially focused pump and Stokes beams, typically in the 700-1,000 nm range, coherently drive a vibrational mode when their frequency difference matches the vibrational resonance. This interaction produces pump depletion (stimulated Raman loss) and Stokes amplification (stimulated Raman gain). In the following analysis, we focus on SRS-loss detection implemented by modulating the Stokes beam between ON and OFF states.

**Indirect photothermal detection by MIP and SRP microscopy**

We next turn from direct vibrational detection to PT detection. PT microscopy measures the thermal response that arises after vibrational relaxation. After vibrational excitation, non-radiative relaxation transfers energy to the surrounding medium, producing a local temperature rise. Because heat diffusion is much slower than vibrational relaxation, which occurs on the picosecond timescales, repeated cycles of excitation and relaxation can lead to thermal accumulation within the probe diffraction-limited volume on nanosecond-to-microsecond timescales, depending on the sample geometry (Fig. 1b). This cumulative temperature rise induces a local RI change, which forms the PT signal.

We first consider MIP microscopy, in which MIR absorption is converted into a photothermal RI change that is read out with a VIS probe beam. In this modality, heat diffusion is a critical determinant of spatial resolution, making the pump-probe timing a key design parameter, as schematically illustrated in Fig. 1c. To clarify this constraint, we consider two nearby absorbers within the MIR excitation region. If heat generated by one absorber diffuses to the location of the other before probing, both absorbers contribute to the measured RI change. The detected response can then no longer be uniquely assigned to either absorber, resulting in signal crosstalk and reduced spatial resolution. This limitation is particularly relevant in MIP because the MIR excitation spot generally spans multiple locations that can be spatially resolved by the VIS readout. When the MIR excitation duration or VIS probe delay extends into the microsecond regime, the thermal profiles from nearby absorbers can spread and overlap before detection, thereby increasing crosstalk.

In contrast, when a single nanosecond MIR pump pulse is followed immediately by a nanosecond VIS probe pulse[15,20], heat diffusion remains spatially confined during the measurement. Under these conditions, crosstalk from nearby absorbers is minimal. Rapid probing also preserves a larger fraction of the locally generated PT response, thereby suppressing signal loss caused by heat diffusion. Nanosecond pump-probe operation is therefore preferred when MIP measurements require both high spatial resolution and high sensitivity. We consequently use this operating condition as the representative MIP case in the following sensitivity comparison.

We next consider SRP microscopy, in which the spatial-resolution constraint imposed by heat diffusion is substantially relaxed. In SRP, tightly focused pump and Stokes beams generate the PT response through SRS excitation, while an independent continuous-wave (CW) VIS probe detects the resulting RI change. Unlike MIP, SRP typically employs a high-repetition-rate picosecond pulse train, usually operating from one to tens of megahertz, which effectively produces burst-pulse PT excitation over microsecond time windows (Fig. 1d). Heat diffusion can be substantial during such a burst, but it produces little crosstalk between nearby absorbers. SRS excitation is confined to the spatial overlap of the focused pump and Stokes beams. This excitation volume typically has a lateral extent of approximately 500 nm, comparable to the spatial resolution of the VIS readout. Nearby absorbers outside this SRS excitation volume are therefore only weakly excited and contribute minimally to the detected PT response. SRP can thus use burst-pulse excitation while largely preserving the spatial resolution defined by the VIS readout.

During the excitation burst, the PT signal increases through thermal accumulation. The signal continues to grow until the system approaches a quasi-steady thermal state, in which heat generation is balanced by diffusive heat loss. Thermal accumulation therefore enhances the PT signal relative to single-pulse excitation.

**Unified framework for shot-noise-limited sensitivity comparison**

Despite their distinct contrast mechanisms, all four modalities detect small changes in an optical signal and can therefore be analyzed within a common contrast-formation framework. In SRS, MIP, and SRP, image contrast is obtained from differential measurements between excitation ON and OFF states, implemented by modulating the Stokes beam in SRS, the MIR beam in MIP, and the pump-plus-Stokes beams in SRP. In MIA, an analogous differential signal arises from the specimen-induced change in transmitted MIR intensity.

The readout schemes can also be described within a unified framework across both point-scanning implementations with photodiode (PD) and wide-field implementations with image-sensor readout. Despite these architectural differences, shot-noise-limited sensitivity can still be compared on a common basis by evaluating the number of detected photons contributing to each image pixel over the total acquisition time. In practice, MIA, SRS, and SRP are usually implemented in the point-scanning format, where a diffraction-limited focus is raster-scanned across the specimen and the signal is detected with a PD. MIP can also be implemented in this way using a CW VIS probe beam. However, because the sensitivity of this point-scanning MIP is strongly affected by heat diffusion, it requires a separate analysis and is not included in the present comparison. High-resolution, high-sensitivity MIP more commonly employs a nanosecond pump-probe scheme described above, and in many practical implementations this scheme is realized in a wide-field format with image-sensor readout over the field of view (FOV)[15,17,20]. We therefore consider wide-field MIP to be a suitable representative MIP implementation for the following sensitivity comparison. A detailed comparison between point-scanning and wide-field MIP sensitivity is provided in Ref. 21.

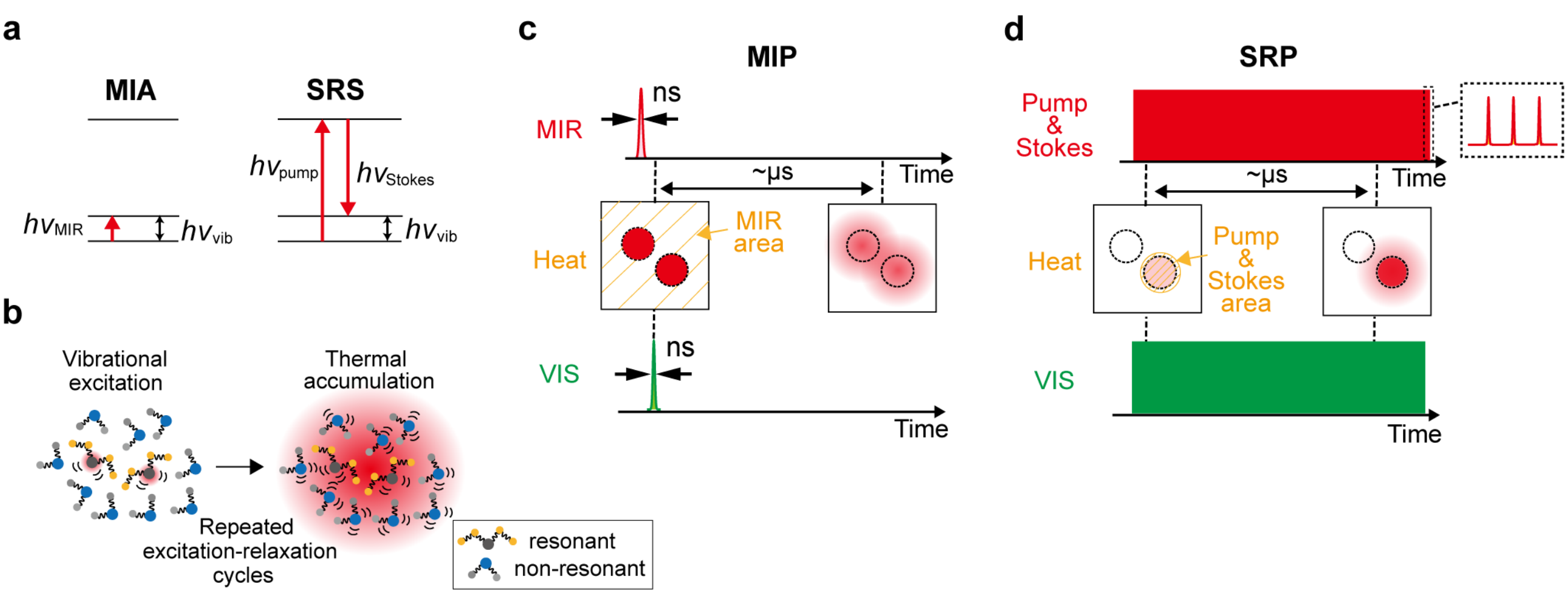

**Fig. 1 Principles of MIA, SRS, MIP, and SRP microscopy. a** Energy-level diagrams of vibrational excitation in MIA and SRS. **b** Repeated excitation-relaxation cycles of vibrationally resonant target molecules (yellow) produce local heating in the surrounding medium (blue), generating a photothermal (PT) response. **c** MIP with nanosecond MIR pump-probe operation. A nanosecond MIR pulse induces a localized PT signal that is read out immediately by a nanosecond VIS probe before substantial heat diffusion occurs. **d** SRP with burst-pulse excitation. High-repetition-rate pump and Stokes pulses generate a PT response over a microsecond time window, while a CW VIS probe detects the resulting RI change. Dotted circles indicate the original object positions, and red shading denotes the PT signal.

## 3. Theory

We establish a unified framework for quantitatively comparing the shot-noise-limited sensitivities of MIA, SRS, MIP, and SRP microscopy. This comparison is both meaningful and theoretically well posed because, in principle, all four modalities can reach the shot-noise limit when optimally designed. On this basis, we derive analytical expressions for the shot-noise-limited sensitivity of each modality. The formulation proceeds in two steps. First, we recast the SRS signal in terms of an effective absorption coefficient, placing SRS on the same footing as linear absorption and thereby enabling direct comparison with MIA. Second, we introduce a PT factor that separates contributions unique to PT readout from those shared with linear absorption. This allows direct comparisons of MIP with MIA and of SRP with SRS. The resulting framework provides a unified basis for quantitative comparison of the shot-noise-limited sensitivities of all four modalities.

### 3.1 Comparison of SRS and MIA.

We compare SRS microscopy with MIA microscopy by formulating their signal-to-noise ratios (SNRs) within a common shot-noise-limited framework. In SRS microscopy, the measured signal is the depletion of pump photons, $\Delta N_{\mathrm{pump}}$, obtained from the difference between the Stokes-ON and Stokes-OFF states. This depletion equals the number of molecules promoted to the vibrationally excited state through the SRS process and can be written as

$$\Delta N_{\mathrm{pump}} = (\sigma_{\mathrm{SRS}} \phi_{\mathrm{stokes}} C_{\mathrm{s}} N_{\mathrm{A}}) l N_{\mathrm{pump}} \equiv (\alpha_{\mathrm{SRS}} l) N_{\mathrm{pump}}, \qquad \text{(Eq. 1)}$$

where $\sigma_{\mathrm{SRS}}$ is the absolute SRS cross-section ($\mathrm{cm}^4\ \mathrm{s\ photon}^{-1}$), as defined in ref 22, $\phi_{\mathrm{stokes}}$ is the Stokes photon flux ($\mathrm{photon\ cm}^{-2}\ \mathrm{s}^{-1}$), $C_{\mathrm{s}}$ is the molecular concentration, $N_{\mathrm{A}}$ is Avogadro's number, $l$ is the effective axial interaction length, and $N_{\mathrm{pump}}$ is the number of incident pump photons. We define $\alpha_{\mathrm{SRS}} \equiv \sigma_{\mathrm{SRS}} \phi_{\mathrm{stokes}} C_{\mathrm{s}} N_{\mathrm{A}}$, which serves as an effective absorption coefficient and places the SRS signal in a form directly comparable to linear absorption.

Let $N_{\mathrm{det_pump}}$ denote the total number of detected pump photons acquired over the Stokes-ON and Stokes-OFF measurements within one pixel dwell time, with $N_{\mathrm{det_pump}}/2$ detected in each measurement. The shot noise associated with the difference between these two measurements is then $\sqrt{2}\sqrt{N_{\mathrm{det_pump}}/2} = \sqrt{N_{\mathrm{det_pump}}}$. By contrast, the SRS-loss signal is $(\alpha_{\mathrm{SRS}} l)(N_{\mathrm{det_pump}}/2)$. The SNR per pixel is therefore

$$SNR_{\mathrm{SRS}} = \frac{(\alpha_{\mathrm{SRS}} l) N_{\mathrm{det_pump}}/2}{\sqrt{N_{\mathrm{det_pump}}}} = \frac{\alpha_{\mathrm{SRS}} l}{2} \sqrt{N_{\mathrm{det_pump}}}. \qquad \text{(Eq. 2)}$$

The SNR for MIA microscopy can be described in an analogous manner. For a differential measurement between the sample-present vs sample-absent states, the shot-noise-limited SNR per pixel is given by

$$SNR_{\mathrm{MIA}} = \frac{\alpha_{\mathrm{MIR}} l}{2} \sqrt{N_{\mathrm{det_MIR}}}, \quad \text{(Eq. 3)}$$

where $N_{\mathrm{det_MIR}}$ is the number of detected MIR photons per image pixel, and $\alpha_{\mathrm{MIR}} \equiv \sigma_{\mathrm{MIR}} C_{\mathrm{s}} N_{\mathrm{A}}$ is the linear absorption coefficient (cm$^{-1}$), with $\sigma_{\mathrm{MIR}}$ denoting the MIA cross-section (cm$^{2}$). If the interaction length $l$ is the same in the two modalities and the detected-photon budget is fixed such that $N_{\mathrm{det_pump}} = N_{\mathrm{det_MIR}}$, then the relative shot-noise-limited sensitivity is determined solely by the comparison between $\alpha_{\mathrm{SRS}}$ and $\alpha_{\mathrm{MIR}}$.

### 3.2 Comparison of MIP and MIA.

We compare MIP microscopy with MIA microscopy by introducing a photothermal (PT) factor that isolates the contribution arising specifically from PT readout. In MIP microscopy, absorption of MIR photons generates the local temperature rise, $\Delta T$, which in turn induces a RI change, $\Delta n$, given by

$$\Delta n = \frac{dn}{dT} \Delta T = \frac{dn}{dT} \frac{(\alpha_{\mathrm{MIR}} l) N_{\mathrm{exc_MIR}} \beta_{\mathrm{MIR}}}{C}, \quad \text{(Eq. 4)}$$

where $N_{\mathrm{exc_MIR}}$ is the number of incident MIR photons per pulse within one image pixel area, $\beta_{\mathrm{MIR}} \equiv h\nu_{\mathrm{MIR}}$ is the MIR photon energy, $C$ is the effective heat capacity of the image voxel, and $dn/dT$ is the thermo-optic coefficient. The heat capacity in water was calculated as $C = \rho C_{\mathrm{p}} (A_{\mathrm{pixel}} l)$, where $\rho$ is the density of water, $C_{\mathrm{p}}$ is the specific heat capacity of water, $A_{\mathrm{pixel}}$ is the image pixel area, and $l$ is the effective axial interaction length.

This RI change is detected through the resulting modulation of the VIS probe beam. As a representative readout scheme, we consider phase-sensitive detection, exemplified by quantitative phase microscopy (QPM). In this case, the signal is the PT-induced phase shift, $\varphi = 2\pi \Delta n l / \lambda_{\mathrm{vis}}$, where $\lambda_{\mathrm{vis}}$ is the wavelength of the VIS probe beam. For a differential measurement between MIR-ON and MIR-OFF states, the shot-noise-limited noise floor is given by the phase precision, $2/(\sqrt{\mu N_{\mathrm{det_vis}}})$, where $N_{\mathrm{det_vis}}$ is the number of detected VIS photons per pixel and $\mu$ is a detection-configuration-dependent sensitivity factor. For a given detector, $\mu$ is defined as the ratio of the maximum VIS illumination that can be delivered to the sample without detector saturation to the corresponding limit in bright-field imaging (see ref. 21 for details). This detection-configuration-dependent factor allows the same formalism to be applied to different VIS microscopy readout schemes, including dark-field (DF) and Zernike phase-contrast microscopy. Typical values of $\mu$ range from approximately 0.25 to 200, corresponding to a $\sqrt{\mu}$ range of 0.5 to 14, which directly determines the VIS-readout sensitivity.

Accordingly, the SNR for MIP microscopy is

$$SNR_{\mathrm{MIP}} = \frac{\alpha_{\mathrm{MIR}} l}{2} \sqrt{N_{\mathrm{det_vis}}} (N_{\mathrm{exc_MIR}} \gamma \sqrt{\mu}), \quad \text{(Eq. 5)}$$

where we define $\gamma \equiv \beta_{\text{MIR}}/C \times dn/dT \times 2\pi l/\lambda_{\text{vis}}$. Under an equal detected-photon budget, $N_{\text{det_MIR}} = N_{\text{det_vis}}$, Eq. 5 can be written as the shot-noise-limited SNR of conventional MIA microscopy multiplied by the PT factor for MIP:

$$SNR_{\text{MIP}} = SNR_{\text{MIA}} \times PT_{\text{MIP}},$$

where

$$PT_{\text{MIP}} = N_{\text{exc_MIR}}\gamma\sqrt{\mu}. \qquad \text{(Eq. 6)}$$

The PT factor therefore quantifies how PT readout modifies the shot-noise-limited sensitivity of MIP relative to MIA. When $PT_{\text{MIP}} > 1$, MIP exceeds conventional MIA in shot-noise-limited sensitivity. Physically, $PT_{\text{MIP}}$ is determined by three quantities: the MIR excitation strength, $N_{\text{exc_MIR}}$; the conversion efficiency $\gamma$, which describes how efficiently absorbed vibrational energy is converted into PT response; and the VIS-readout sensitivity, $\sqrt{\mu}$. These dependencies highlight the need to jointly optimize MIR excitation and VIS readout sensitivity.

### 3.3 Comparison of SRP and SRS.

We compare SRP microscopy with SRS microscopy by taking into account the additional PT contribution arising from burst-pulse thermal accumulation. Unlike MIP, SRP contains an extra PT term because high-repetition-rate SRS excitation deposits vibrational energy over many successive pulses. When excitation-relaxation cycles occur faster than heat can diffuse out of the focal region, thermal energy accumulates locally. As repeated heating continues, the system approaches a quasi-steady thermal state in which heat input is balanced by diffusive loss. In this regime, the temperature increase can be approximated by the steady-state heat conduction relation

$$\Delta T_{\text{s}} \sim Q/2\pi\kappa L, \qquad \text{(Eq. 7)}$$

where $Q$ is the heat input per unit time (W), $\kappa$ is the thermal conductivity (W $m^{-1}$ $K^{-1}$), and $L$ is the characteristic diameter of the heated region (m)[23]. For SRP, the heat input is

$$Q = (\alpha_{\text{SRS}} l) N_{\text{pump/pulse}} f_{\text{rep}} \beta_{\text{vib}}, \qquad \text{(Eq. 8)}$$

where $N_{\text{pump/pulse}}$ is the number of pump photons per pulse, $f_{\text{rep}}$ is the laser repetition rate, and $\beta_{\text{vib}} \equiv h(\nu_{\text{pump}} - \nu_{\text{stokes}}) = h\nu_{\text{vib}}$ is the vibrational energy. Substituting Eq. 8 into Eq. 7 yields the steady-state temperature rise

$$\Delta T_{\text{s}} = \frac{(\alpha_{\text{SRS}} l) N_{\text{pump/pulse}} f_{\text{rep}} \beta_{\text{vib}}}{2\pi\kappa L} \equiv \frac{(\alpha_{\text{SRS}} l) N_{\text{pump/pulse}} \beta_{\text{vib}}}{C} N_{\text{accum}}, \qquad \text{(Eq. 9)}$$

where the second form rewrites $\Delta T_{\text{s}}$ as a single-pulse temperature rise, analogous to Eq. 4, multiplied by an effective number of pulses, $N_{\text{accum}}$, that contribute to the quasi-steady thermal state. Here, $C$ is the heat capacity associated with the image voxel, as defined in Section 3.2. Equating the two forms in Eq. 9 gives

$$N_{\mathrm{accum}} = \frac{f_{\mathrm{rep}} C}{2\pi\kappa L}. \qquad \text{(Eq. 10)}$$

When a target is comparable to or larger than the SRS excitation volume, the characteristic size $L$ is determined by the SRS excitation geometry rather than by the physical target size, with $L \propto V_{\mathrm{SRS}}{}^{1/3}$ (see "SRS excitation geometry" in Methods). Because $C$ is proportional to the measurement voxel volume, which is set by the SRS excitation volume, $C \propto V_{\mathrm{SRS}} \propto L^3$. Equation 10 therefore gives $N_{\mathrm{accum}} \propto C/L \propto L^2$. This scaling reflects the fact that larger heat sources dissipate heat more slowly and therefore allow stronger thermal accumulation, whereas smaller heat sources lose heat more rapidly and suppress accumulation.

Comparison of the temperature increases in Eq. 9 (SRP) and Eq. 4 (MIP) shows that SRP replaces $N_{\mathrm{exc_MIR}}$ with $N_{\mathrm{pump/pulse}}$ and introduces an additional multiplicative factor, $N_{\mathrm{accum}}$, arising from burst-pulse thermal accumulation. The PT factor for SRP is therefore

$$PT_{\mathrm{SRP}} = N_{\mathrm{pump/pulse}} \gamma \sqrt{\mu} N_{\mathrm{accum}}, \qquad \text{(Eq. 11)}$$

where $\gamma\sqrt{\mu}$ is unchanged from MIP under the same VIS detection scheme, provided that the same vibrational transition is probed ($\beta_{\mathrm{vib}} = \beta_{\mathrm{MIR}}$).

**4. Numerical analysis.**

We quantify the theoretical sensitivities of SRS, MIP, and SRP by substituting experimentally realistic parameters into Eqs. 2, 5, 6 and 11. Representative SRS ($\sigma_{\mathrm{SRS}}$) and MIA ($\sigma_{\mathrm{MIR}}$) cross-sections are listed in Table 1[4,22,24]. To compare Raman-based and MIR-based modalities on a consistent basis, we selected vibrational bands that satisfy two criteria: they have relatively large $\sigma_{\mathrm{SRS}}$ or $\sigma_{\mathrm{MIR}}$ values in the corresponding off-resonant Raman or MIR absorption processes, and they are relevant to bioimaging through either exogenous probes or endogenous biomolecules. For the SRS and SRP analysis, we selected three representative off-resonant Raman modes. These were the alkyne C≡C stretch of EdU, a widely used exogenous probe in the cell-silent region; the conjugated/aromatic C=C stretch in retinol, which is relevant to endogenous unsaturated and aromatic biomolecules; and the C–H stretch of DMSO, used here as a C–H vibration relevant to endogenous lipids. For the MIP analysis, we consider three representative MIA bands: amide I, which arises primarily from the peptide-backbone C=O stretch of proteins; the ester C=O stretch, which is commonly found in lipids; and the phosphate P=O stretch, which is commonly found in nucleic acids.

We then estimate the effective absorption coefficients per 1 M bond concentration, $\alpha_{\mathrm{SRS}}$ and $\alpha_{\mathrm{MIR}}$, for the representative vibrational bands introduced above. To evaluate $\alpha_{\mathrm{SRS}}$, we consider a widely used near-IR implementation of SRS or SRP with a Stokes wavelength of 1,030 nm. Because $\alpha_{\mathrm{SRS}}$ scales linearly with the Stokes power, we adopt a representative condition of 50 mW Stokes excitation (1.25 nJ per pulse at 80 MHz; 50% duty cycle), which is typical of highly sensitive cellular imaging[13]. The estimated values of $\alpha_{\mathrm{SRS}}$ and $\alpha_{\mathrm{MIR}}$ are summarized in Table 1. Under this experimentally relevant condition, $\alpha_{\mathrm{MIR}}$ remains substantially larger than $\alpha_{\mathrm{SRS}}$ even when compared with the strong off-resonant Raman bands considered here. Specifically, $\alpha_{\mathrm{MIR}}$ is 660-1,800 cm$^{-1}$, whereas $\alpha_{\mathrm{SRS}}$ is 2.2-22.2 cm$^{-1}$ for the strong off-resonant Raman bands, corresponding to an approximately 30–800-fold difference. These results highlight the intrinsically higher sensitivity of MIA-based microscopy.

**Table 1 Representative SRS and MIR cross-sections and effective absorption coefficients.** Representative SRS/MIR cross-sections ($\sigma_{\mathrm{SRS}}$, $\sigma_{\mathrm{MIR}}$), and corresponding effective absorption coefficients ($\alpha_{\mathrm{SRS}}$, $\alpha_{\mathrm{MIR}}$) were calculated for a bond concentration of 1 M. For SRS, the Stokes beam power was assumed to be 50 mW, corresponding to 1.25 nJ per pulse at 80 MHz with a 50% duty cycle at 1,030 nm. The Stokes photon flux $\phi_{\mathrm{stokes}}$ was calculated as described in "Stokes photon flux" in Methods. GM: 1 x 10$^{-50}$ cm$^4$ s photon$^{-1}$.

| Modality | Molecule / group | Vibrational mode | Wavenumber (cm$^{-1}$) | Cross-section | Effective absorption coefficient (cm$^{-1}$) |
|---|---|---|---|---|---|
| SRS / SRP | EdU | C≡C stretch | 2,125 | 10 GM | 22.2 |
| SRS / SRP | Retinol | C=C stretch | 1,595 | 4 GM | 8.9 |
| SRS / SRP | DMSO | C-H stretch | 2,913 | 1 GM | 2.2 |
| MIA / MIP | Protein / peptide bond | C=O stretch (amide I) | 1,620-1,650 | 3.0 × 10$^{-18}$ cm$^2$ | 1,800 |
| MIA / MIP | Ester carbonyl | C=O stretch (ester) | 1,690-1,770 | 1.1 × 10$^{-18}$ cm$^2$ | 660 |
| MIA / MIP | Phosphate | P=O stretch | 1,200-1,300 | 1.9 × 10$^{-18}$ cm$^2$ | 1,140 |

We next evaluate the PT factors, $PT_{\mathrm{SRP}}$ and $PT_{\mathrm{MIP}}$, in a water environment under the conditions considered here. The geometrical and thermal parameters required for these calculations are summarized in the caption of Table 2, and the calculation details are provided in Methods. To ensure a consistent comparison, we set the VIS-readout sensitivity to a representative value of $\sqrt{\mu}$ = 1 for both modalities[21]. For SRP, we assume a representative pump power of 30 mW, corresponding to 0.375 nJ per pulse at 80 MHz[25] as $PT_{\mathrm{SRP}}$ scales linearly with pump power. We also assume that a target is comparable to or larger than the SRS excitation volume. Under this condition, the characteristic diameter of the heated region in SRP, $L$, is determined by the SRS excitation geometry, resulting in $L$~ 682 nm (see "SRS excitation geometry" in Methods). For MIP, we assume a wide-field illumination over 61 μm × 61 μm with representative MIR powers of 0.5 mW and 2.5 mW, corresponding to 0.5 μJ and 2.5 μJ per pulse at 1 kHz, respectively[15]. The rationale for selecting these excitation powers is mentioned in Discussion.

The resulting values of $PT_{\mathrm{SRP}}$ and $PT_{\mathrm{MIP}}$ are summarized in Table 2. For SRP, burst-pulse excitation leads to substantial thermal accumulation, with $N_{\mathrm{accum}} = 26$, which increases $PT_{\mathrm{SRP}}$ to 2.6. Without this accumulation, namely for $N_{\mathrm{accum}} = 1$, $PT_{\mathrm{SRP}}$ would be only 0.10. Burst-pulse thermal accumulation is therefore required for SRP to exceed SRS in shot-noise-limited sensitivity, corresponding to $PT_{\mathrm{SRP}} > 1$. By contrast, $PT_{\mathrm{MIP}}$ remains one to two orders of magnitude smaller than the accumulated $PT_{\mathrm{SRP}}$. Under single-pulse conditions, however, the two PT factors are similar, with $PT_{\mathrm{SRP}}$=0.10 and $PT_{\mathrm{MIP}}$ =0.04-0.21. This comparison indicates that the PT-factor advantage of SRP over MIP arises specifically from burst-pulse accumulation rather than from an intrinsically larger single-pulse PT response.

In both modalities, the PT factor can also be enhanced through the VIS-detection term $\sqrt{\mu}$. This term depends on the readout scheme and can vary by as much as an order of magnitude, from ~1 to ~10. Highly sensitive VIS detection methods, such as dark-field (DF)[26] and interferometric scattering (iSCAT)[27] microscopy, can raise $\sqrt{\mu}$ by about one order of magnitude. Under such favorable detection conditions, not only $PT_{\mathrm{SRP}}$ but also $PT_{\mathrm{MIP}}$ can exceed unity, indicating that MIP, like SRP, can surpass direct MIA in shot-noise-limited sensitivity. This increase in $\sqrt{\mu}$, however, is generally achievable only for submicron-sized targets because of the operating principles of DF and iSCAT microscopy[27].

**Table 2 Representative PT factors for SRP and MIP microscopy**. Representative SRP and MIP photothermal factors ($PT_{\mathrm{SRP}}$, $PT_{\mathrm{MIP}}$) were calculated for a target vibrational band at 1,660 cm$^{-1}$. The VIS-detection sensitivity factor was set to $\sqrt{\mu}$ = 1. For SRP, the pump wavelength was set to 880 nm, the pump pulse energy was 0.375 nJ at 80 MHz, and the characteristic heated region diameter was set to $L = 682$ nm based on the SRS excitation geometry. The thermal conductivity $\kappa$ of water was set to 0.6 W m$^{-1}$ K$^{-1}$ [28]. For MIP, MIR excitation at 6 μm was delivered over an 61 μm × 61 μm field of view, with pulse energies of 0.5 and 2.5 μJ at 1 kHz. Calculation details for heat capacity $C$ and conversion efficiency $\gamma$ are provided in "Heat capacity" and "Conversion efficiency" in Methods.

| Modality | Excitation condition | $N_{\mathrm{accum}}$ | PT factor |
|---|---|---|---|
| SRP | Pump 30 mW<br>(0.375 nJ at 80 MHz) | 26 | 2.6 |
| MIP | MIR 0.5 mW<br>(0.5 μJ at 1 kHz) | N/A | 0.04 |
| MIP | MIR 2.5 mW<br>(2.5 μJ at 1 kHz) | N/A | 0.21 |

Finally, we benchmark the SNRs of SRS, SRP, and MIP for a target vibrational band at 1,660 cm$^{-1}$ and a bond concentration of 10 mM. The comparison assumes a 150 × 150-pixel FOV, corresponding to 61 μm × 61 μm, acquired in 1 s. The pixel pitch is matched to the lateral SRS excitation size, approximately 404 nm, as defined in "SRS excitation geometry" in Methods. We use $\sigma_{\mathrm{SRS}}$ = 5 GM and $\sigma_{\mathrm{MIR}}$ = 2 × 10$^{-18}$ cm$^2$ as representative values based on Table 1. The SNR calculation uses the Stokes photon flux defined for Table 1 and the PT factors summarized in Table 2. For clarity, the excitation powers, per-pulse energies, and duty cycles used here are stated explicitly below.

For MIP, the entire field is illuminated with 2.5 mW MIR light, corresponding to 2.5 μJ per pulse at 1 kHz. For SRS, the excitation powers are 30 mW for the pump beam, corresponding to 0.375 nJ per pulse at 80 MHz with a 100% duty cycle, and 50 mW for the Stokes beam, corresponding to 1.25 nJ per pulse at 80 MHz with a 50% duty cycle. For SRP, the pump and Stokes beams are burst-gated at 50% duty cycle to generate PT-ON and PT-OFF states while preserving the same per-pulse energies used in SRS, corresponding to a 15 mW average pump power at 0.375 nJ per pulse and a 50 mW average Stokes power at 1.25 nJ per pulse. For a fair comparison of shot-noise-limited SNR, a CW VIS probe is used for SRP detection with its power set to match the detected-photon budget of the SRS pump beam, such that $N_{\mathrm{det_vis}} = N_{\mathrm{det_pump}}$. Complete calculation parameters are provided in the associated public repositories[29].

Under the benchmark conditions described above, the SNRs of SRS and SRP are 16.4 and 42.7, respectively. These values correspond to an approximately 2.6-fold higher SNR for SRP than for SRS in the ideal shot-noise-limited comparison. This result indicates that, under matched pulse-energy and detection conditions, the intrinsic SNR gain of SRP over SRS is limited to a few-fold increase in aqueous cellular environments, rather than an order-of-magnitude enhancement. In practical implementations, this theoretical advantage may be further affected by nonideal beam overlap among the pump, Stokes and VIS probe beams, as well as by laser intensity noise at the lower effective modulation frequencies used for burst-gated SRP detection.

For MIP, although $\alpha_{\mathrm{MIR}}$ is about two orders of magnitude larger than $\alpha_{SRS}$, this intrinsic advantage is strongly offset by the small PT factor ($PT_{\mathrm{MIP}}$ = 0.21). Consequently, even with detected-photon budgets matched to those of SRS and SRP, the ideal MIP SNR remains at 364. In practice, however, this ideal value is further reduced because the full-well capacity of the image sensor limits the detected-photon budget available in wide-field MIP. Even with high-full-well-capacity sensors[15], the total optical power deliverable to the detector is limited to ~1 mW, reducing the achievable SNR to 47.7 (see "Photon budget of image sensors" in Methods). These results suggest that, under current wide-field MIP detector constraints, the practically achievable MIP SNR is around 50, comparable to that of SRP under the present benchmark conditions.

**Table 3 Benchmark comparison of SNRs in SRS, SRP and MIP microscopy.** SNRs were calculated for a target vibrational band at 1,660 $cm^{-1}$ and a bond concentration of 10 mM. The comparison assumes a 150 × 150-pixel FOV, corresponding to 61 μm × 61 μm, acquired in 1 s. For MIP, the MIR illumination area was set to 61 μm × 61 μm. The VIS-detection sensitivity factor was set to $\sqrt{\mu}$ = 1. The photon-budget-matched MIP case uses the same detected-photon budget as SRS and SRP, whereas the sensor-limited MIP case uses the detected-photon budget allowed by the full-well capacity of the image sensor.

| Modality | Cross-section | Excitation condition | Detected photons ($N_{\mathrm{det_pump}}$, $N_{\mathrm{det_vis}}$) | SNR |
|---|---|---|---|---|
| SRS | 5 GM | Pump (880 nm): 30 mW, 0.375 nJ $pulse^{-1}$, 80 MHz, 100% duty<br><br>Stokes (1,030 nm): 50 mW, 1.25 nJ $pulse^{-1}$, 80 MHz, 50% duty | $5.9 \times 10^{12}$ | 16.4 |
| SRP | 5 GM | Pump (880 nm): 15 mW, 0.375 nJ $pulse^{-1}$, 80 MHz, 50% duty<br><br>Stokes (1,030 nm): 50 mW, 1.25 nJ $pulse^{-1}$, 80 MHz, 50% duty | $5.9 \times 10^{12}$ | 42.7 |
| MIP (photon-budget-matched) | $2 \times 10^{-18}$ $cm^2$ | MIR (6 μm): 2.5 mW, 2.5 μJ $pulse^{-1}$, 1 kHz | $5.9 \times 10^{12}$ | 364 |
| MIP (sensor-limited) | $2 \times 10^{-18}$ $cm^2$ | MIR (6 μm): 2.5 mW, 2.5 μJ $pulse^{-1}$, 1 kHz | $1.0 \times 10^{11}$ | 47.7 |

**Discussion**

**Validation of theoretical SNR with experimental data.**

To assess whether our theoretical SNR estimate is experimentally reasonable, we applied our SNR calculation framework to experimental parameters from previously reported SRS[30] and wide-field MIP[15] studies. The calculated SNRs were consistent with the reported experimental evaluations, as described in "Validation of theoretical SNR with experimental data in SRS" and "Validation of theoretical SNR with experimental data in MIP" in Methods.

We also compared our estimate with a previous SRP experiment that reported an approximately 500-fold signal enhancement in DMSO[18]. At first glance, this result appears inconsistent with the 2.6-fold SRP-to-SRS improvement predicted by our theoretical calculation. However, this apparent discrepancy can be largely explained by two differences between our theoretical comparison and the previous experimental study: the excitation conditions and the sample properties.

First, our comparison keeps the per-pulse energies used for SRS unchanged when evaluating SRP. In contrast, the previous study kept the average power identical between SRS and SRP while reducing the SRP duty cycle, thereby increasing the SRP per-pulse energy by approximately one order of magnitude. Second, our calculation assumes a water environment, whereas the previous study used DMSO. Relative to water, DMSO gives an approximately fivefold larger $\gamma$, owing to its larger $dn/dT$, and an approximately threefold larger $N_{\mathrm{accum}}$, owing to its lower thermal conductivity $\kappa$. These material-dependent differences together lead to an ~15-fold enhancement in the PT factor. These two considerations account for the major part of the previously reported several-hundred-fold enhancement.

**Rationale for selecting excitation powers.**

The practical sensitivity of SRS, SRP and MIP is ultimately constrained by the excitation energy that can be delivered to biological samples without inducing damage. The limiting mechanism differs between Raman-based and MIR-based modalities. In SRS and SRP, the signal increases with the near-infrared pump and Stokes photon fluxes, but the usable pulse energy is limited by photodamage associated with tightly focused ultrashort pulses, including nonlinear electronic excitation, oxidative stress and other sample-dependent responses[31-33]. In MIP, MIR absorption is a low-photon-energy linear process and therefore avoids multiphoton electronic transitions, but the pulse energy is instead limited by transient heating and thermal accumulation, particularly in aqueous environments.

For the SRS and SRP benchmark, we used 0.375 nJ per pump pulse and 1.25 nJ per Stokes pulse during the ON state at 80 MHz. These values are within the range reported for biological SRS imaging. Cellular lipid SRS imaging with an 80 MHz picoEmerald system has used 0.125–0.375 nJ pump pulses and 0.125–0.625 nJ Stokes pulses[34], while EdU imaging in HeLa cells has used 0.31 nJ pump and 0.63 nJ Stokes pulses in a 2 ps mode, and up to 0.75 nJ pump and 1.50 nJ Stokes pulses in a 14 ps mode[33]. Higher-power live-cell and in vivo SRS studies have also used low-nJ pulse energies[35]. Because the damage threshold depends not only on pulse energy but also on pulse duration, repetition rate, dwell time, wavelength, focusing geometry and sample type, this excitation condition should be regarded as a practical high-sensitivity benchmark rather than a theoretical upper limit.

For the MIP benchmark, we select MIP pulse energies of 0.5 and 2.5 μJ per pulse at 1 kHz over a 61 μm × 61 μm wide-field illumination area. These values were chosen by estimating the transient temperature rise caused by water absorption using Eq. 4[36]. To maintain biological viability, we allowed a transient temperature increase of up to 7 K[37]. For pure water at 56 M, this criterion corresponds to a maximum molar extinction coefficient of 17 $M^{-1}$ $cm^{-1}$ for a 0.5 μJ pulse and 3.4 $M^{-1}$ $cm^{-1}$ for a 2.5 μJ pulse. The 0.5 μJ condition is therefore compatible with strongly absorbing water bands, such as the amide I region around 1,600-1,700 $cm^{-1}$, where the molar extinction coefficient is ~15 $M^{-1}$ $cm^{-1}$. By contrast, the 2.5 μJ condition is more appropriate for moderately absorbing regions, such as 1,750-2,990 $cm^{-1}$, where the molar extinction coefficient is ~3 $M^{-1}$ $cm^{-1}$, or 1,000-1,550 $cm^{-1}$, where it is ~5 $M^{-1}$ $cm^{-1}$. It is also compatible with $D_2O$-based imaging, for which the amide I region has a molar extinction coefficient of ~2 $M^{-1}$ $cm^{-1}$.

**Can burst-pulse excitation improve MIP sensitivity?**

The MIP benchmark used in the sensitivity comparison assumes nanosecond single-pulse excitation to avoid the spatial-resolution degradation caused by heat diffusion during burst-pulse excitation. This choice does not preclude the use of microsecond burst-pulse excitation in MIP at the expense of spatial resolution. However, burst-pulse excitation does not inherently improve MIP sensitivity.

In principle, if each pulse in a burst could retain the MIR energy used for nanosecond single-pulse excitation, the total MIR energy delivered during the burst would increase with the effective number of accumulated pulses, $N_{\text{accum}}$. Such operation could provide an accumulation benefit analogous to that considered for SRP. In MIP, however, this condition is generally not permissible when the damage or viability limit is determined by the allowable temperature rise in water. In MIP, intracellular water absorbs MIR light throughout the MIR-illuminated volume, producing a heated volume that is typically much larger than the localized volume heated in SRP. On the microsecond timescale of a burst, heat diffusion from this extended volume is limited, allowing the temperature rise to accumulate across successive pulses.

Under these conditions, the temperature rise at the end of a burst is determined primarily by the total MIR energy delivered during the burst, $E_{\text{burst}} = N_{\text{accum}} E_{\text{pulse}}$, rather than by how that energy is distributed among individual pulses. A burst that delivers the same total MIR energy as a nanosecond single pulse is consequently expected to produce a comparable final temperature rise in water. When water-induced heating sets the damage or viability limit, increasing $N_{\text{accum}}$ cannot increase the allowable total MIR energy. Burst excitation consequently provides no intrinsic gain in MIP sensitivity.

SRP may benefit from burst-pulse accumulation through a different mechanism. During picosecond SRS excitation, the usable excitation energy may be limited, at least in part, by nonlinear optical damage. Distributing a given excitation energy across multiple pulses can reduce the peak intensity relevant to nonlinear damage while allowing the PT response to accumulate. This regime may permit an increase in $N_{\text{accum}}$ and thereby improve sensitivity. The key distinction between MIP and SRP is therefore not simply how excitation energy is delivered in time, but which physical process limits the usable excitation energy. In the water-heating-limited regime considered here, MIP is constrained by linear MIR absorption in water, whereas SRP may be limited by nonlinear damage associated with tightly focused picosecond Raman excitation. A clearer understanding of these damage pathways will be required to establish the ultimate sensitivity limits of both methods.

## Conclusion

We quantitatively compared the sensitivities of SRS, MIP, and SRP microscopy by introducing two key quantities: the effective absorption coefficient and the PT factor. Direct MIA benefits from absorption cross-sections that are much larger than the corresponding Raman scattering cross-sections. Under excitation conditions relevant to biological imaging, the MIR absorption coefficient, $\alpha_{\mathrm{MIR}}$, for polar vibrational modes such as the amide I (C=O), ester C=O, and phosphate P=O stretches typically exceeds the effective SRS absorption coefficient, $\alpha_{\mathrm{SRS}}$, for strong off-resonant Raman modes, including C–H stretches, alkyne C≡C stretches, and conjugated or aromatic C=C stretches, by approximately two orders of magnitude.

In MIP microscopy, however, this intrinsic advantage is partly offset by the small PT factor ($PT_{\mathrm{MIP}} \approx 0.1$). Consequently, the net shot-noise-limited sensitivity gain over SRS is typically limited to several-fold and, at most, about one order of magnitude. In SRP microscopy, by contrast, the smaller $\alpha_{\mathrm{SRS}}$ is compensated by a much larger PT factor ($PT_{\mathrm{SRP}} \approx 3$), enabled by burst-pulse excitation that accumulates heat on microsecond timescales. Under ideal shot-noise-limited conditions with matched per-pulse excitation energy, SRP provides sensitivities that are several-fold higher than those of SRS in water and can approach those of MIP.

## Methods

### Stokes photon flux.

Following ref. 22, we calculated the Stokes photon flux from the average Stokes power at the sample plane. The peak Stokes photon flux is given by

$$\phi_{\mathrm{stokes}} = \frac{P_{\mathrm{stokes}}}{\beta_{\mathrm{stokes}} A_{\mathrm{stokes}} f_{\mathrm{rep}} \tau_{\mathrm{pulse}}},$$

where $P_{\mathrm{stokes}}$ is the average Stokes power at the sample during the ON period, $\beta_{\mathrm{stokes}} = h\nu_{\mathrm{stokes}}$ is the Stokes photon energy, and $A_{\mathrm{stokes}}$ is the effective focal area of the Stokes beam, $f_{\mathrm{rep}}$ is the repetition rate, and $\tau_{\mathrm{pulse}}$ is the pulse duration. The focal area is calculated as

$$A_{stokes} = \pi\left(\sqrt{2} w_{xy_stokes}\right)^2,$$

where $w_{\mathrm{xy_stokes}}$ is the lateral $1/e$ radius of the Stokes focal spot, and $\sqrt{2} w_{\mathrm{xy_stokes}}$ corresponds to its $1/e^2$ radius. The lateral $1/e$ radius is given by

$$w_{\mathrm{xy_stokes}} = \frac{0.325\lambda_{\mathrm{stokes}}}{\sqrt{2}NA^{0.91}},$$

where $\lambda_{\mathrm{stokes}}$ is the Stokes wavelength and $NA$ is numerical aperture of the objective. Using $P_{\mathrm{stokes}}$ = 100 mW, $f_{\mathrm{rep}}$= 80 MHz, and $\tau_{\mathrm{pulse}}$= 5 ps, $\lambda_{\mathrm{stokes}}$= 1,030 nm, and $NA$=1.0, we obtain $\phi_{\mathrm{stokes}}$= 3.70 x $10^{29}$ photon $cm^{-2}$ $s^{-1}$.

**SRS excitation geometry.**

Following ref. 22, the SRS excitation geometry was estimated from the pump focal geometry using multiphoton excitation-volume approximation. The lateral and axial $1/e$ radii of the pump focus are given by

$$w_{\mathrm{xy_pump}} = \frac{0.325\lambda_{\mathrm{pump}}}{\sqrt{2}NA^{0.91}},$$

and

$$w_{\mathrm{z_pump}} = \frac{0.532\,\lambda_{\mathrm{pump}}}{\sqrt{2}\,(n_{\mathrm{m}}-\sqrt{n_{\mathrm{m}}{}^{2}-NA^{2}})},$$

where $\lambda_{\mathrm{pump}}$ is the pump wavelength and $n_m$ is the RI of the surrounding medium. The lateral $1/e$ diameter was used to define the image pixel pitch in the SNR evaluation. The effective axial interaction length, $l$, was defined as the axial full width at half maximum (FWHM), $2\sqrt{ln2}\,w_{\mathrm{z_pump}}$[38]. Using $NA$ =1.0, $\lambda_{\mathrm{pump}}$= 880 nm, and $n_{\mathrm{m}}$= 1.33, we obtain an image pixel pitch of 404 nm and an effective axial interaction length of 1.22 µm.

The effective SRS excitation volume was calculated as

$$V_{\mathrm{SRS}} = \pi^{\frac{3}{2}}\, w_{\mathrm{xy_pump}}{}^{2}\, w_{\mathrm{z_pump}}.$$

This gives $V_{\mathrm{SRS}}$= 1.66 × $10^{-1}$ $\mu m^3$. When this volume is converted into an equivalent sphere, the corresponding effective diameter is

$$L = \left(\frac{6V_{\mathrm{SRS}}}{\pi}\right)^{1/3},$$

giving $L$ = 682 nm.

**Heat capacity.**

The effective lumped heat capacity of the heated voxel in water was calculated as

$$C = \rho C_{\mathrm{p}}(A_{\mathrm{pixel}}l),$$

where $\rho$ is the density of water, $C_{\mathrm{p}}$ is the specific heat capacity of water, $A_{\mathrm{pixel}}$ is the image pixel area, and $l$ is the effective axial interaction length. The pixel area was defined as

$$A_{\text{pixel}} = \left(2w_{\text{xy_pump}}\right)^2,$$

using the lateral $1/e$ diameter of the effective pump focus. With $\rho$ = 1,000 kg m$^{-3}$, $C_p$ = 4,180 J kg$^{-1}$ K$^{-1}$, and the optical parameters defined in "SRS excitation geometry", we obtain $C$ = 8.32 ×10$^{-13}$ J K$^{-1}$.

**Conversion efficiency**.

The conversion efficiency $\gamma$ describes how efficiently absorbed vibrational energy is converted into a VIS-detectable PT phase response. It is defined as

$$\gamma \equiv \frac{\beta_{\text{MIR}}}{C} \times \frac{dn}{dT} \times \frac{2\pi l}{\lambda_{\text{vis}}},$$

where $\beta_{\text{MIR}}$ is the MIR photon energy, $C$ is the lumped heat capacity calculated above, $dn/dT$ is thermo-optic coefficient, $l$ is the effective axial interaction length, and $\lambda_{\text{vis}}$ is the VIS probe wavelength. Using $\lambda_{\text{vis}}$= 500 nm, $dn/dT$= 1 x 10$^{-4}$ K$^{-1}$, $C$ = 8.32 ×10$^{-13}$ J K$^{-1}$, and $\beta_{\text{MIR}}$ = 3.30 x 10$^{-20}$ J for a wavelength of 6 µm, we obtain $\gamma$ = 6.08 ×10$^{-11}$. Because SRP and MIR absorption excite the same vibrational transition, we set $\beta_{\text{vib}} = \beta_{\text{MIR}}$ and use the same value of $\gamma$ for SRP.

**Validation of theoretical SNR with experimental data in SRS.**

To validate our SRS sensitivity calculation, we compared its predictions with previously reported measurements[30]. In that study, SRS imaging was performed using pump and Stokes wavelengths of 812 nm and 1,064 nm, respectively, with 1.6-ps pulses and an objective NA of 1.2. The average powers at the sample plane were 10 mW for the pump beam, corresponding to 0.125 nJ per pulse at 80 MHz with 100% duty cycle, and 72 mW for the Stokes beam, corresponding to 1.8 nJ per pulse at 80 MHz with a 50% duty cycle. The reported sensitivity limit was 19 mM for DMSO at a pixel dwell time of 2 µs, corresponding to an imaging rate of 22.2 frames per second for a 150 × 150-pixel image. After normalizing this measurement to a 1-s acquisition time, the experimental sensitivity limit corresponds to approximately 4 mM.

Using an SRS cross-section for DMSO of 1 GM[24] and the reported experimental parameters, our theoretical calculation yielded a sensitivity limit of ~1.7 mM. This value is reasonably consistent with the experimentally inferred limit, supporting the validity of the theoretical SNR framework for SRS.

**Calculation of photon budget for image sensors.**

In wide-field MIP, the detected-photon budget available for each image pixel is limited by the full-well capacity of the image sensor and can be calculated as

$$N_{\text{det_vis}} = \frac{N_{\text{full-well}} N_{\text{sens_pixel}} N_{\text{frame}}}{N_{\text{im_pixel}}},$$

where $N_{\text{full-well}}$ is the full-well capacity per sensor pixel, $N_{\text{sens_pixel}}$ is the total number of sensor pixels, $N_{\text{frame}}$ is the number of acquired frames, and $N_{\text{im_pixel}}$ is the number of image pixels. In our calculation, we used parameters from Adimec Q-2HFW camera: $N_{\text{full-well}} = 2 \times 10^6$, $N_{\text{sens_pixel}} = 1{,}440 \times 1{,}440$, $N_{\text{frame}} = 550$ over a 1 s acquisition, and $N_{\text{im_pixel}} = 150 \times 150$.

**Validation of theoretical SNR with experimental data in MIP.**

To validate our MIP sensitivity calculation, we compared its predictions with previously reported wide-field MIP measurements based on off-axis digital holography[15]. This system imaged intracellular lipid droplets through the $CH_2$ stretching band near 2,925 $cm^{-1}$ at 50 frames per second over 245 × 245 pixels. Under MIR illumination of ~6.5 µJ per pulse over an 87 µm × 87 µm area, lipid droplets with a diameter of ~2 µm were detected with an SNR of ~90. The $CH_2$-bond concentration in lipid droplets was approximated as ~5 M from the lipid mass concentration of ~0.07 g/ml[39], using a representative neutral-lipid approximation with 20 $CH_2$ bonds per molecule and a molecular weight of 282 g $mol^{-1}$, consistent with the neutral-lipid-rich composition of lipid droplets[40]. Because the reported experiment did not use the full photon budget available from a high full-well-capacity image sensor, the measured SNR could be increased by a factor of ~4.5 under detector-limited optimal conditions. After accounting for this potential 4.5-fold SNR improvement and normalizing the measurement to a 1-s acquisition with 150 × 150 pixels, the experimentally inferred sensitivity limit is estimated to be ~1.1 mM.

The corresponding theoretical sensitivity was calculated using a representative MIR absorption cross-section of $\sigma_{\text{MIR}} = 5 \times 10^{-19}$ $cm^2$ for the $CH_2$ asymmetric stretching band. This value is of the same order as peak cross-sections estimated from reported integrated band strengths and spectral linewidths of saturated aliphatic hydrocarbons[41]. For example, reported integrated absorption strengths of 5.5-7.4 × $10^{-18}$ cm $group^{-1}$ per $CH_2$ group and FWHM linewidths of 22-28 $cm^{-1}$ correspond to peak cross-sections of ~2-3.4 × $10^{-19}$ $cm^2$. Using the experimental parameters described above and a VIS-readout sensitivity factor of $\sqrt{\mu}$= 0.5 for off-axis digital holography, our framework predicts a sensitivity limit of 0.8 mM for a 1-s acquisition with 150 × 150 pixels. This value is in reasonable agreement with the experimentally inferred limit, supporting the validity of the theoretical SNR framework for wide-field MIP.

**Data availability**

A complete set of calculation parameters is available in the associated public repositorie[29].

**Acknowledgments**

This work was financially supported by Japan Society for the Promotion of Science (23H00273, 25H01386, T.I.), JST FOREST Program (JPMJFR236C, T.I.), RIKEN TRIP initiative (T.I.). We thank Zicong Xu for critically reading the manuscript.

**Author contributions**

K.T. performed theoretical analysis. K.T. and T.I. discussed the results and wrote the manuscript.

## Competing interests

The authors declare no competing interests.